\documentstyle[sprocl,epsf,rotate]{article}
\input{psfig}
\newcommand{\IM}{{\rm Im}}
\newcommand{\vcb}{|V_{cb}|}
\newcommand{\vtd}{|V_{td}|}
\newcommand{\vub}{|V_{ub}/V_{cb}|}
\newcommand{\vts}{|V_{ts}|}

\def\epe{\varepsilon'/\varepsilon}

\newcommand{\mt}{m_{\rm t}}
\newcommand{\mtb}{\overline{m}_{\rm t}}

\newcommand{\mc}{m_{\rm c}}

\newcommand{\mb}{m_{\rm b}}
\newcommand{\mw}{M_{\rm W}}
\newcommand{\mz}{M_{\rm Z}}
\newcommand{\gev}{\, {\rm GeV}}
\newcommand{\mev}{\, {\rm MeV}}

\newcommand{\bea}{\begin{eqnarray}}
\newcommand{\eea}{\end{eqnarray}}
\newcommand{\bd}{\begin{displaymath}}
\newcommand{\ed}{\end{displaymath}}

\newcommand{\beq}{\begin{equation}}
\newcommand{\eeq}{\end{equation}}
\newcommand{\be}{\begin{equation}}
\newcommand{\ee}{\end{equation}}
\newcommand{\ord}{{\cal O}}

\def\kpn{K^+\rightarrow\pi^+\nu\bar\nu}
\def\klpn{K_{\rm L}\rightarrow\pi^0\nu\bar\nu}

\begin{document}
\begin{flushright}
 TUM-HEP-299/97 \\
 hep-ph/9711217 \\
 October 1997
\end{flushright}
\vskip1truecm
\centerline{\LARGE\bf  CKM Matrix: Present and Future
   \footnote[1]{\noindent Invited talk given at
 the "Symposium on Heavy Flavours", Santa Barbara, July 7 - July 11,
1997, to
appear in the proceedings.}}
  \vskip1truecm
\centerline{\Large\bf Andrzej J. Buras}
\bigskip
\centerline{\sl Technische Universit{\"a}t M{\"u}nchen, Physik
Department}
\centerline{\sl D-85748 Garching, Germany}
\vskip1truecm
\thispagestyle{empty}
\centerline{\bf Abstract}

We review the present status of the CKM matrix and we offer some visions
of its future. After a brief presentation of the theoretical framework
for
weak decays we discuss the following topics:
i) CKM matrix from tree level decays,
ii) Standard analysis  of the unitarity triangle,
iii) CKM matrix from rare and CP violating K- and B-decays,
iv) CKM matrix from CP violation in two-body B-decays.
In particular we compare the CKM potentials of the standard analysis of
the unitarity triangle, of the very clean
$ K \to \pi\nu\bar\nu $ decays and of CP asymmetries in B-decays.

\vfill\eject

\title{CKM MATRIX: PRESENT AND FUTURE}

\author{ ANDRZEJ J. BURAS }

\address{Technische Universit\"at M\"unchen, Physik Department,\\
D-85748 Garching, Germany}


\maketitle\abstracts{
We review the present status of the CKM matrix and we offer some visions
of its future. After a brief presentation of the theoretical framework for
weak decays we discuss the following topics:
i) CKM matrix from tree level decays,
ii) Standard analysis  of the unitarity triangle,
iii) CKM matrix from rare and CP violating K- and B-decays,
iv) CKM matrix from CP violation in two-body B-decays.
In particular we compare the CKM potentials of the standard analysis of
the unitarity triangle, of the very clean
$ K \to \pi\nu\bar\nu $ decays and of CP asymmetries in B-decays.}


\section{Introduction}
\subsection{ CKM Matrix}
An important target of particle physics is the determination
of the unitary $3\times 3$ Cabibbo-Kobayashi-Maskawa
matrix \cite{CAB,KM}, $\hat V_{CKM}$, which parametrizes the charged
current interactions of  quarks.
CP violation in the Standard Model is supposed to arise
from a single phase in this matrix.
 The standard parametrization of the CKM matrix
 \cite{PDG} is given in terms of
$c_{ij}=\cos\theta_{ij}$, $s_{ij}=\sin\theta_{ij}$ and the phase $\delta$
with $i$ and $j$
being generation labels ($i,j=1,2,3$).
$s_{13}$ and $s_{23}$ turn out to be small numbers:
$\ord(10^{-3})$ and ${\cal O}(10^{-2})$,
respectively. Consequently to an excellent accuracy
the four independent parameters to be determined experimentally are:
\begin{equation}\label{2.73}
s_{12}=| V_{us}|, \quad s_{13}=| V_{ub}|, \quad s_{23}=|
V_{cb}|, \quad \delta
\end{equation}

On the other hand, in the phenomenological applications,
it is customary these days to work
with the CKM-matrix expressed in
terms of four Wolfenstein parameters
\cite{WO} $(\lambda,A,\varrho,\eta)$
with $\lambda=\mid V_{us}\mid=0.22 $ playing the role of an expansion
parameter and $\eta$
representing the CP violating phase:
\begin{equation}\label{2.75}
\hat V_{CKM}=
\left(\begin{array}{ccc}
1-{\lambda^2\over 2}&\lambda&A\lambda^3(\varrho-i\eta)\\ -\lambda&
1-{\lambda^2\over 2}&A\lambda^2\\ A\lambda^3(1-\varrho-i\eta)&-A\lambda^2&
1\end{array}\right)
+O(\lambda^4)
\end{equation}
The Wolfenstein parametrization is certainly more transparent than the
standard parametrization. However, if one requires sufficient level
of accuracy, the higher order terms in $\lambda$
have to be included in phenomenological
applications. An efficient and systematic way to achieve this without
the lost of transparency
is to {\it define} the parameters
$(\lambda, A, \varrho, \eta)$ through \cite{BLO}
\begin{equation}\label{wop}
s_{12}\equiv\lambda \qquad s_{23}\equiv A \lambda^2 \qquad
s_{13} e^{-i\delta}\equiv A \lambda^3 (\varrho-i \eta)
\end{equation}
Making this change of variables in the standard parametrization,
we find the CKM matrix as a function of
$(\lambda,A,\varrho,\eta)$ which, in contrast to (\ref{2.75})
 satisfies unitarity exactly!
Expanding next in powers of $\lambda$ we recover the
matrix in (\ref{2.75}) and in addition find explicit corrections of
$\ord(\lambda^4)$ and higher order terms.

The definition of $(\lambda,A,\varrho,\eta)$ given in (\ref{wop})
is useful because it allows to improve the accuracy of the
original Wolfenstein parametrization in an elegant manner. In
particular
\begin{equation}\label{CKM1}
V_{us}=\lambda \qquad V_{cb}=A\lambda^2
\end{equation}
\begin{equation}\label{CKM2}
V_{ub}=A\lambda^3(\varrho-i\eta)
\qquad
V_{td}=A\lambda^3(1-\bar\varrho-i\bar\eta)
\end{equation}
\begin{equation}\label{2.83d}
 V_{ts}= -A\lambda^2+\frac{1}{2}A(1-2 \varrho)\lambda^4
-i\eta A \lambda^4
\end{equation}
where
\begin{equation}\label{3}
\bar\varrho=\varrho (1-\frac{\lambda^2}{2})
\qquad
\bar\eta=\eta (1-\frac{\lambda^2}{2})
\end{equation}
turn out \cite{BLO} to be excellent approximations to the
exact expressions.

The advantage of this generalization of the Wolfenstein parametrization
over other generalizations found in the literature is the absence of
relevant corrections to $V_{us}$, $V_{cb}$ and $V_{ub}$.
Corrections to $V_{us}$ and $V_{cb}$ appear only at $\ord(\lambda^7)$ and
$\ord(\lambda^8)$, respectively. $V_{ub}$ as given above is valid to
all orders.
Simultaneously the simple modification of
$V_{td}$ relative to the one in (\ref{2.75})
 allows a simple generalization of the unitarity
triangle beyond the leading order in $\lambda$.

\begin{figure}[hbt]
\vspace{0.10in}
\centerline{
\epsfysize=1.5in
\epsffile{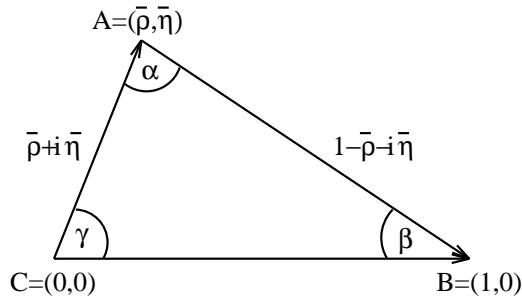}
}
\vspace{0.08in}
\caption[]{
Unitarity Triangle.
\label{fig:utriangle}}
\end{figure}

The unitarity
triangle is obtained by using the unitarity relation
\begin{equation}\label{2.87h}
V_{ud}V_{ub}^* + V_{cd}V_{cb}^* + V_{td}V_{tb}^* =0,
\end{equation}
rescaling it by $\mid V_{cd}V_{cb}^\ast\mid=A \lambda^3$ and depicting
the result in the complex $(\bar\rho,\bar\eta)$ plane as shown
in fig. 1. The lenghts CB, CA and BA are equal respectively to 1,
\begin{equation}\label{2.94a}
R_b \equiv  \sqrt{\bar\varrho^2 +\bar\eta^2}
= (1-\frac{\lambda^2}{2})\frac{1}{\lambda}
\left| \frac{V_{ub}}{V_{cb}} \right|
\qquad
{\rm and}
\qquad
R_t \equiv \sqrt{(1-\bar\varrho)^2 +\bar\eta^2}
=\frac{1}{\lambda} \left| \frac{V_{td}}{V_{cb}} \right|.
\end{equation}

The triangle in fig. 1, $\mid V_{us}\mid$ and $\mid V_{cb}\mid$
give the full description of the CKM matrix.
Looking at the expressions for $R_b$ and $R_t$ we observe that within
the Standard Model the measurements of four CP
{\it conserving } decays sensitive to $|V_{us}|$, $\vcb$,
$| V_{ub}| $ and $ |V_{td}|$ can tell us whether CP violation
($\bar\eta \not= 0$) is predicted in the Standard Model.
This is a very remarkable property of
the Kobayashi-Maskawa picture of CP violation: quark mixing and CP violation
are closely related to each other. This property is often used to determine
the angles of the unitarity triangle without the study of CP violating
quantities.

There is of course the very important question whether the KM picture
of CP violation is correct and more generally whether the Standard
Model offers a correct description of weak decays of hadrons. In order
to answer these important questions it is essential to calculate as
many branching ratios as possible, measure them experimentally and
check if they all can be described by the same set of the parameters
$(\lambda,A,\varrho,\eta)$. In the language of the unitarity triangle
this means that the various curves in the $(\bar\varrho,\bar\eta)$ plane
extracted from different decays should cross each other at a single point
as shown in fig. 2.
Moreover the angles $(\alpha,\beta,\gamma)$ in the
resulting triangle should agree with those extracted one day from
CP-asymmetries in B-decays. More about this below.

\begin{figure}[hbt]
\vspace{0.10in}
\centerline{
\epsfysize=4.0in
\rotate[r]{
\epsffile{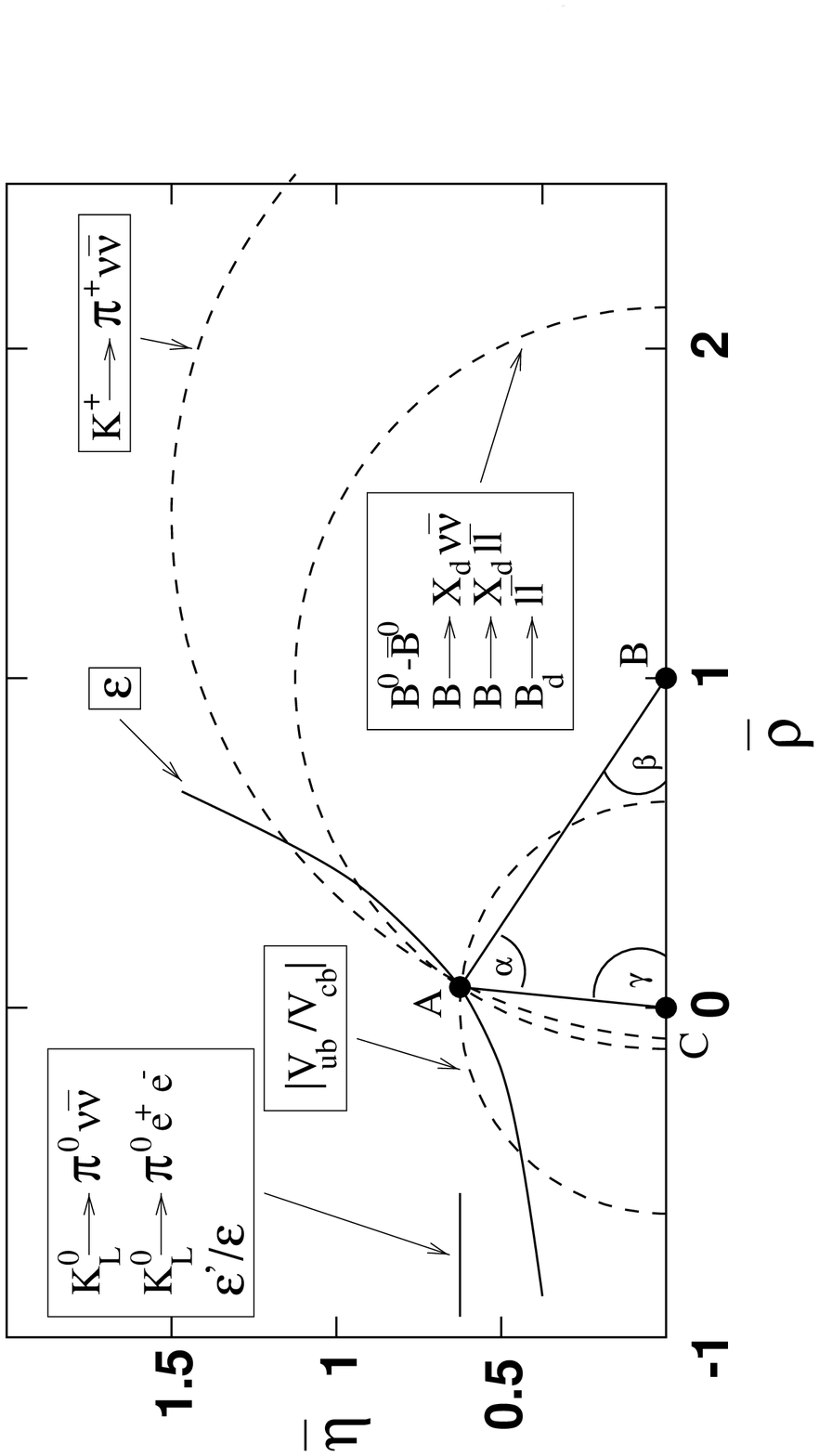}
}}
\vspace{0.08in}
\caption[]{
The ideal Unitarity Triangle. For artistic reasons the value of
$\bar\eta$ has been chosen to be higher than the fitted central value
$\bar\eta\approx 0.35.$
\label{fig:2011}}
\end{figure}

Since the CKM matrix is only a parametrization of quark mixing and
of CP violation and does not offer the explanation of these two
very important phenomena, many physicists hope that a new physics
while providing a dynamical origin of quark mixing and CP violation will
also change the picture given in fig. 2. That is, the different curves
based on the Standard Model expressions, will not cross each other
at a single point
and the angles $(\alpha,\beta,\gamma)$
extracted one day from
CP-asymmetries in B-decays will disagree with the ones determined from
rare K and B decays.

Clearly the plot in fig. 2 is highly idealized because in order
to extract such nice curves from various decays one needs perfect
experiments and perfect theory.
One of the goals of this review is to identify those decays
for which at least the theory is under control. For such decays,
if they can be measured with a sufficient precision, the curves
in fig. 2 are
 not unrealistic.
 In order to understand this we have to discuss briefly the present
theoretical framework.

\subsection{Theoretical Framework}
The basic problem in the extraction of CKM parameters from meson decays
is related to strong
interactions. Although due to
the smallness
of the effective QCD coupling at short distances, the gluonic
contributions at scales ${\cal O} (\mw, \mz, \mt)$ can be calculated
within the perturbative framework, the fact that
 mesons are $ q\bar q$ bound states forces us to consider  QCD at
long distances as well.
 Here we have to rely on existing non-perturbative
methods, which are not yet very powerful
at present.

The separation of the short and long distance contributions to a given
amplitude is achieved by using
the powerful method of the Operator Product Expansion
(OPE) combined with the renormalization group
approach.
Explicitly
the amplitude for an {\it exclusive} decay $M\to F$
is written as
\begin{equation}\label{OPE}
 A(M \to F) = \frac{G_F}{\sqrt 2} {\rm V_{CKM}} \sum_i
   C_i (\mu) \langle F \mid Q_i (\mu) \mid M \rangle
\end{equation}
where ${\rm V_{CKM}}$ denotes the relevant CKM factor.
The scale $ \mu $ separates the physics contributions in the ``short
distance'' contributions (corresponding to scales higher than $\mu $)
contained in the Wison coefficients $ C_i(\mu) $ and
the ``long distance'' contributions
(scales lower than $ \mu $) contained in
$\langle F \mid Q_i (\mu) \mid M \rangle $.
Here $Q_i$ are
local operators generated by QCD and electroweak interactions
 which govern ``effectively'' the decays in question.
The $\mu$ dependence of $ C_i(\mu) $
is governed by the renormalization group equations.
It must be canceled by the one present in $\langle  Q_i (\mu)\rangle $
so that the resulting physical amplitudes do not depend on $\mu$.
Generally this cancellation involves many
operators due to the operator mixing under renormalization.
The list of the operators originating in the tree diagrams,
in one-loop diagrams of fig. 3 and in QCD corrections to them,
as well as
technical details of the calculations of
$C_i(\mu)$ can be found in \cite{BBL}.

The use of the renormalization group
is  necessary in order to sum up large logarithms
 $ \log \mw/\mu $ which appear for $ \mu= {\cal O}(1-2\gev) $.
 In the so-called leading
logarithmic approximation (LO) terms $ (\alpha_s\log \mw/\mu)^n $ are summed.
The next-to-leading logarithmic correction (NLO) to this result involves
summation of terms $ (\alpha_s)^n (\log \mw/\mu)^{n-1} $ and so on.
This hierarchic structure gives the renormalization group improved
perturbation theory.

\begin{figure}[hbt]
\vspace{0.10in}
\centerline{
\epsfysize=3in
\rotate[r]{
\epsffile{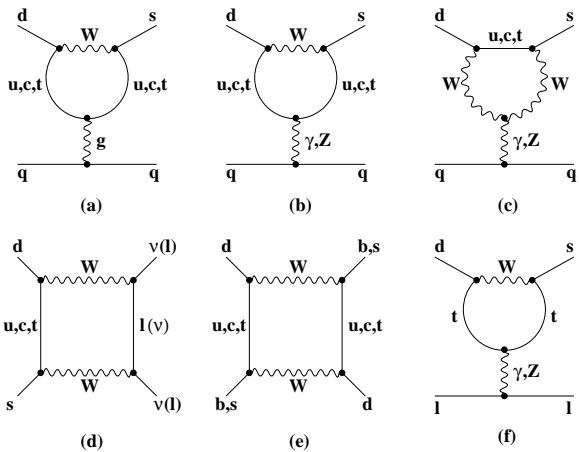}
}}
\vspace{0.08in}
\caption[]{
Typical Penguin and Box Diagrams.
\label{fig:fdia}}
\end{figure}

The rather formal expression for the decay amplitudes given in
(\ref{OPE}) can always be cast in the form:
\begin{equation}\label{PBEE}
A(M\to F)=\sum_i B_i {\rm V^i_{CKM}} \eta^{i}_{QCD} F_i(\mt,\mc)
\end{equation}
which is more useful for phenomenology. In writing (\ref{PBEE})
we have generalized (\ref{OPE}) to include several CKM factors.
$F_i(m_t,m_c)$, the Inami-Lim functions \cite{IL},
 result from the evaluation of loop diagrams with
internal top and charm exchanges (see fig. 3) and may also depend
solely on $\mt$ or $\mc$. In the case of new physics they
depend  on masses of new particles such as charginos, stops and charged
Higgs scalars as well as new couplings.
The factors $\eta^{i}_{QCD}$ summarize
the QCD corrections which can be calculated by formal methods
mentioned above. Finally $B_i$ stand for nonperturbative factors
related to the hadronic matrix elements of the contributing
operators: the main theoretical uncertainty in the whole enterprise.
In leptonic and semi-leptonic decays for which only the matrix elements
of weak currents are needed,
the non-perturbative $B$-factors can fortunately be determined from
leading tree level decays reducing
or removing the non-perturbative uncertainty. In non-leptonic
decays this is generally not possible and we have to rely on
existing non-perturbative methods. A well known example of a
$B_i$-factor is the renormalization group invariant parameter
$B_K$  defined by
\begin{equation}\label{bk}
B_K=B_K(\mu)\left[\alpha_s(\mu)\right]^{-2/9}
\qquad
\langle \bar K^{0}\mid
(\bar s d)_{V-A} (\bar s d)_{V-A}\mid K^{0}\rangle=
\frac{8}{3} B_K(\mu)F_K^2 m_K^2
\end{equation}

In order to achieve sufficient precision, $C_i(\mu)$
or equvalently $\eta^{i}_{QCD}\equiv\eta_i$ have
to include NLO corrections in the renormalization group
improved perturbation
theory. In particular, unphysical left-over $\mu$-dependences in
decay amplitudes, resulting from the truncation of the perturbative
series, are considerably reduced by including NLO corrections.
These corrections are known by now
for the most important and interesting decays and are
reviewed in \cite{BBL}.

In the case of decaying B-mesons, the approach discussed here can
be generalized to inclusive decays where it is known under the
name of {\it Heavy Quark} ($1/\mb)$ {\it Expansions}.
The leading term in these expansions corresponds to the spectator
model corrected for short distance QCD effects. HQE are discussed
by Thomas Mannel in these proceedings.

\section{CKM Matrix from Tree Level Decays and Unitarity}
\subsection{The seven elements of the CKM Matrix}
The seven elements of the CKM matrix which have been determined in
tree level decays without using unitarity are given as follows:
\begin{equation}\label{vud}
|V_{ud}| = 0.9736 \pm 0.0010\
\qquad
|V_{us}| = \lambda =  0.2205 \pm 0.0018
\end{equation}
as given in \cite{PDG} and
\begin{equation}\label{CHALK}
|V_{ud}|=0.9740\pm 0.0005
\end{equation}
 obtained subsequently
at the Chalk River Laboratory
\cite{Hagberg}.

\begin{equation}
|V_{cd}|=0.224\pm0.016
\qquad
|V_{cs}| = 1.01 \pm 0.18
\end{equation}
taken from \cite{PDG},
\begin{equation}\label{v23}
\vcb=(40\pm 3) \cdot 10^{-3}
\qquad
|V_{ub}|=(3.2\pm0.8)\cdot 10^{-3}
\end{equation}
discussed below
and
\be
|V_{tb}|=0.99\pm0.15
\ee
as obtained from CDF \cite{Narain}.

We do not have any tree level results for $\vts$ and $\vtd$.
Because of very small branching ratios for such
tree level top quark decays we will most likely be able to determine
these elements
only in loop induced decays discussed in subsequent sections.

\subsection{Including Unitarity of CKM}

Setting $\lambda=0.22$, scanning $\vcb$ and $|V_{ub}|$ in
the ranges (\ref{v23})  and $\cos\delta$ in
the range $-1\leq \cos\delta\leq 1$, we find \cite{BF97}:
\begin{equation}\label{uni1}
4.5\cdot 10^{-3}\leq \vtd \leq 13.7\cdot 10^{-3}\,,
\qquad
0.0353\leq \vts \leq 0.0429
\end{equation}
and
\begin{equation}\label{uni2}
0.9991\leq |V_{tb}| \leq 0.9993,
\qquad
0.9736\leq |V_{cs}| \leq 0.9750.
\end{equation}

{}From (\ref{uni1}) we observe
that the unitarity of the CKM matrix requires approximate equality of
$\vts$ and $\vcb$:
$0.954\leq |V_{ts}|/\vcb \leq 0.997$
which is evident if one compares (\ref{CKM1}) with (\ref{2.83d}).
The determination of $\vtd$
will be  improved in the next section by using the
constraints from $B^0_d-\bar B^0_d$-mixing and CP violation in
the $K$-meson system.

Using (\ref{CHALK}),  $|V_{us}|$ and $|V_{ub}|$  one
finds \cite{Hagberg}
\begin{equation}
|V_{ud}|^2 +|V_{us}|^2 +|V_{ub}|^2 =0.9972\pm0.0013,
\end{equation}
where the contribution of $|V_{ub}|^2$ is negligible.
Thus the departure from the unitarity constraint
is by at least two standard deviations. The simplest
solution to this ``unitarity problem'' would be to double
the error in $|V_{ud}|$ or to increase its value. Since
the neutron decay data give, in contrast to $0^+\to 0^+$ superallowed
beta decays, values
for the unitarity sum higher than unity \cite{Hagberg}, such a shift
is certainly possible. Clearly the current status
of the $|V_{ud}|$ determinations, in spite of small errors quoted
above, is unsatisfactory at present.
Next, as stressed by Ben Nefkens at this symposium, the experimental error
in the determination of
$|V_{us}|$ in (\ref{vud}) is underestimated and a new improved experiment
with the
aim of determining this very important element to a high accuracy
is desirable.
It is clear that further investigations of $|V_{ud}|$ and $|V_{us}|$
 should be made before one could conclude
that the failure to meet the unitarity constraint signals
some physics beyond the Standard Model.

\subsection{More on $\vcb$ and $|V_{ub}|$}
\subsubsection{$\vcb$ from Inclusive Decays}
The determination of $\vcb$ from inclusive decays uses
\be\label{inclw}
\Gamma(B\to X_c l \nu_l)=\frac{G_F^2 m_b^5}{192\pi^3} \vcb^2
f_1(\mb,\mc,\lambda_1,\lambda_2,\alpha_s)
\ee
where $\lambda_i$ are the well known non-perturbative parameters
entering the $1/m_b^2$ corrections to the spectator model
result.

The most recent analysis \cite{BSU97}
updates the 95 analysis of
Shifman, Uraltsev and Vainshtein \cite{SUV} with the result:
\be\label{vcbincl}
\vcb_{incl}=41.0 \left[\frac{Br(B\to X_cl\nu)}{0.104}\right]^{1/2}
\left[\frac{1.6 ps}{\tau_B}\right]^{1/2} \cdot 10^{-3}
\cdot \kappa
\ee
where
\be\label{777}
\kappa=
 1 \pm 0.015_{pert}\pm 0.01_{m_b} \pm 0.012_{\lambda_1}\pm 0.012_{rest}.
\ee
The last error is an educated guess for the uncertainties coming
from assuming hadron-quark duality as well as other small corrections
not shown explicitly in (\ref{777}). The main progress this year
has been the reduction of the perturbative uncertainty by a factor
of approximately three through the two-loop calculations of
Czarnecki and Melnikov \cite{CZMI}.
The rather small error due to the uncertainty in
$m_b$ is related to the fact that the leading $m_b^5$ dependence
in (\ref{inclw}) is substantially reduced through the additional
phase space effects collected in $f_1$. This uncertainty is probably
somewhat underestimated.

Taking the estimate (\ref{vcbincl}) with $\tau_B=(1.60\pm0.03) ps$
and $Br(B \to X_c l \nu)= 10.4 \pm 0.4\%$ one arrives at:
\be\label{inclf}
\vcb_{incl}=\left(\begin{array}{cc}
41.0 & [BSUV] \\
38.8 & [BBB] \end{array} \pm 2.0_{th}\pm 0.9_{exp}\right)
\cdot 10^{-3}
\ee
where the second estimate is based on the 95-analysis of
Ball, Benecke and Braun \cite{Braun} with appropriate update in
$\tau_B$ and $Br(B \to X_c l \nu)$. \cite{BSU97} and \cite{Braun}
agree on the size of the theoretical error.
\subsubsection{$\vcb$ from Exclusive Decays}
The exclusive determination of $\vcb$ uses the recoil spectrum
of $D^*$ in $B \to D^* l \bar\nu$:
\be
\frac{d\Gamma(B\to D^*l\bar\nu_l)}{d\omega}= f_2(m_B,m_{D^*},\omega)
\vcb^2 {\cal F}^2(\omega)
\ee
where $f_2$ collects kinematical factors and
${\cal F} (\omega)$ is the relevant formfactor which in the absence
of power and short distance QCD corrections reduces to the Isgur-Wise
function. ${\cal F} (\omega)$ can be calculated reasonably at the
no-recoil point:
\be
{\cal F} (1)= 1+\ord (\alpha_s)+ \ord(\frac{1}{m_c^2})
+\ord(\frac{1}{m_c m_b})+\ord(\frac{1}{m_b^2})
\ee
The perturbative short distance corrections are fully under control
after the complete two-loop analysis of Czarnecki \cite{CZ96}. On the other
hand there is some dispute between Neubert \cite{Neubert} and the authors
in \cite{BSU97} on the uncertainty due to $\ord(1/m^2)$
corrections:
\be
{\cal F} (1)=0.91 \pm 0.01_{pert} \pm
\left[\begin{array}{cc}
0.025 &[N] \\
0.050 & [BSUV] \end{array}\right]_{1/m^2}=
0.91 \pm
\left[\begin{array}{cc}
0.03 &[N] \\
0.06 & [BSUV] \end{array}\right].
\ee
Using Gibbons 97 - Analysis:
\be
{\cal F} (1)\vcb = (35.1\pm 2.5) \cdot 10^{-3}
\ee
one arrives at
\be\label{exclf}
\vcb_{excl} = \left( 38.6 \pm \left[\begin{array}{cc}
1.2 & [N] \\
2.4 & [BSUV] \end{array} \right]_{th}
\pm 2.7_{exp} \right) \cdot 10^{-3}
\ee
I have no intention to make an analysis which somehow combines
the inclusive and exclusive determinations. However, looking
at (\ref{inclf}) and (\ref{exclf}), it appears that
$\vcb=0.040\pm0.003$ as given in (\ref{v23})
is a good summary of the present value for $\vcb$. Clearly values
as $\vcb=0.039\pm0.003$ or $\vcb=0.039\pm0.004$ would also be fine.
On the other hand an error on $\vcb$ as low as 0.002 \cite{Neubert}
appears to me  difficult to defend at present.
\subsubsection{$|V_{ub}|/\vcb$ }
The present status and future prospects for $|V_{ub}|$ have been
discussed by Ball \cite{Ball97}, Gibbons and Flynn at this symposium and
consequently I will
be only very brief on this topic.
The main results at present come from the end-point lepton energy
spectrum in $ B\to X_u e \bar\nu_e$ which are affected by some
model dependence. This model dependence can be somewhat reduced
by using the exclusive decays $B\to \rho(\pi) l \nu_l$
\cite{CLEOU,Gibbons}. Yet the
final result using this method contains still a $\pm 25\%$ error
as seen in (\ref{v23}) and in
\begin{equation}\label{v13}
\frac{|V_{ub}|}{\vcb}=0.08\pm0.02
\end{equation}
With the improved data and improved formfactors coming from lattice
calculations and light-cone sum rules \cite{Ball97},
this error could be possibly decreased to $\pm 10\%$.

On the other hand there is a hope that the study of
the hadronic energy spectrum or hadron invariant mass spectrum
in $ B\to X_u e \bar\nu_e$ \cite{GR96} could allow a measurement
of $|V_{ub}|$ with an error of $\pm 10\%$. Furthermore as suggested
by Uraltsev \cite{URAL},
the ultimate measurement of $|V_{ub}|$ could be obtained
from the inclusive semileptonic $b \to u$ rate by means of:

\be\label{vubincl}
|V_{ub}|=0.00458\cdot \left[\frac{Br(B\to X_ul\nu)}{0.002}\right]^{1/2}
\left[\frac{1.6 ps}{\tau_B}\right]^{1/2} \cdot
[1\pm 0.025_{pert} \pm 0.03_{m_b}].
\ee
In the case of very good data this would allow a $\pm 5\%$ measurement
of $|V_{ub}|$. The measurement of the inclusive semileptonic $b \to u$
rate is very difficult, however, and it will take some time before
this idea could be efficiently realized. In addition the view
that this way of measuring $|V_{ub}|$ is theoretically very clean
is not shared by everybody.

Finally one should recall that the measurement of the leptonic
branching ratio for $B^+ \to l^+ \nu_l$ ($l=e,\mu,\tau$) determines the
product $|V_{ub}| F_B$ with essentially no theoretical uncertainties.
Thus provided such branching ratio can be measured with respectable
accuracy and $F_B$ can be calculated with high precision one day,
one would have another independent measurement of $|V_{ub}|$.

\section{Standard Analysis of the Unitarity Triangle}

\subsection{Basic Formulae}

The standard analysis of the unitarity triangle
proceeds in five
steps:

{\bf Step 1:}

{}From  $b\to c$ transition in inclusive and exclusive B meson decays
one finds $\vcb$ and consequently the scale of UT:
\begin{equation}
\vcb\quad =>\quad\lambda \vcb= \lambda^3 A
\end{equation}

{\bf Step 2:}

{}From  $b\to u$ transition in inclusive and exclusive B meson decays
one finds $\vub$ and consequently the side $CA=R_b$ of UT:
\begin{equation}
\vub \quad=> \quad R_b=4.44 \cdot \left| \frac{V_{ub}}{V_{cb}} \right|
\end{equation}

{\bf Step 3:}

{}From the observed indirect CP violation in $K \to \pi\pi$ described
experimentally by the parameter $\varepsilon_K$ and theoretically
by the imaginary part of the relevant box diagram in fig. 3 one
derives the constraint:
\begin{equation}\label{100}
\bar\eta \left[(1-\bar\varrho) A^2 \eta_2 S(x_t)
+ P_0(\varepsilon) \right] A^2 B_K = 0.226
\qquad
S(x_t)=0.784 \cdot x_t^{0.76}.
\end{equation}
Equation (\ref{100}) specifies
a hyperbola in the $(\bar \varrho, \bar\eta)$
plane.
Here
$x_t={\mt^2}/{\mw^2}$,
$\eta_2$
is the QCD factor in the box diagrams with two top quark exchanges and
$P_0(\varepsilon)=0.31\pm0.05$, dependent on QCD factors $\eta_1$ and
$\eta_3$,  summarizes the contributions
of box diagrams with two charm quark exchanges and the mixed
charm-top exchanges. The error in $P_0(\varepsilon)$ is dominated by
the uncertainties in $\eta_3$ and $\mc$. Since the $P_0(\varepsilon)$
term contributes only $25\%$ to (\ref{100}) these uncertainties
constitute only to a few percent uncertainty in the constraint
(\ref{100}).
The NLO values of the QCD factors
are:
$\eta_1=1.38\pm 0.20$ \cite{HNa},
$\eta_2=0.57\pm 0.01$ \cite{BJW} and
  $\eta_3=0.47\pm0.04$ \cite{HNb}.
The quoted errors reflect the remaining theoretical uncertainties due to
$\mu$-dependences and $\Lambda_{\overline{MS}}$ .

$B_K$
is the non-perturbative parameter defined in (\ref{bk}).
The review of its values in various non-perturbative approaches
can be found in \cite{BF97}. I only mention here new results not
included there.
The  most accurate lattice results for $B_K$
come from JLQCD $B_K=0.87\pm0.06$ \cite{JLQCD}.
Similar result
has been published in \cite{GKS} this year.
On the other hand a recent analysis in
the chiral quark model gives surprisingly a value as high as
$B_K=1.3\pm 0.2$
\cite{BERT97}. In our numerical analysis presented
below we will use, as in \cite{BF97}, $B_K=0.75\pm 0.15$
which is in the ball park of various lattice estimates and
$B_K=0.70\pm 0.10$ from
the $1/N$ approach \cite{BBG0}.

Since $m_t$ and the relevant QCD factors are rather precisely known,
the main uncertainties in the constraint (\ref{100}) reside in
$B_K$ and to some extent in $A^4$ which multiplies the leading term.

{\bf Step 4:}

{}From the observed $B^0_d-\bar B^0_d$ mixing described experimentally
by the mass difference $\Delta M_d$ or by the
mixing parameter $x_d=\Delta M_d/\Gamma_B$
and theoretically by the relevant box diagram of fig. 3
the side $BA=R_t$ of the UT can be determined:

\begin{equation}\label{106}
 R_t= \frac{1}{\lambda}\frac{|V_{td}|}{\vcb} = 1.0 \cdot
\left[\frac{|V_{td}|}{8.8\cdot 10^{-3}} \right]
\left[ \frac{0.040}{\vcb} \right]
\end{equation}
with
\begin{equation}\label{VT}
\vtd=
8.8\cdot 10^{-3}\left[
\frac{200\mev}{\sqrt{B_{B_d}}F_{B_d}}\right]
\left[\frac{170~GeV}{\mtb(\mt)} \right]^{0.76}
\left[\frac{\Delta M_d}{0.50/{\rm ps}} \right ]^{0.5}
\sqrt{\frac{0.55}{\eta_B}}.
\end{equation}

Here $\eta_B$ is the QCD factor analogous to $\eta_2$ and given
by $\eta_B=0.55\pm0.01$ \cite{BJW}. Next
$F_{B_d}$ is the B-meson decay constant and $B_{B_d}$
denotes a non-perturbative
parameter analogous to $B_K$.

There is a vast literature on the lattice calculations of $F_{B_d}$
and $B_{B_d}$.
The world averages given by Flynn \cite{Flynn} and Bernard \cite{Bernard}
can be summarized by:
$F_{B_d}=175\pm 25~MeV$ and
$B_{B_d}=1.31\pm 0.03$.
This result for $F_{B_d}$ is compatible with the results obtained using
QCD sum rules \cite{QCDSF}.
In our numerical analysis we will use
$F_{B_d}\sqrt{B_{B_d}}=200\pm 40~MeV$. The experimental situation on
$\Delta M_d$ taken from Gibbons \cite{Gibbons}
 is given in table 1.

Since $\mt$, $\Delta M_d$ and $\eta_B$ are already rather precisely
known, the main uncertainty in the determination of $\vtd$ from
$B_d^0-\bar B_d^0$ mixing comes from $F_{B_d}\sqrt{B_{B_d}}$.
Note that $R_t$ suffers from additional uncertainty in $\vcb$,
which is absent in the determination of $\vtd$ this way.

{\bf Step 5:}

{}The measurement of $B^0_s-\bar B^0_s$ mixing parametrized by $\Delta M_s$
together with $\Delta M_d$  allows to determine $R_t$ in a different
way. Setting $\Delta M^{max}_d= 0.482/ps$ and
$|V_{ts}/V_{cb}|^{max}=0.993$ (see table 1) I find
a useful formula:
\begin{equation}\label{107b}
(R_t)_{max} = 1.0 \cdot \xi \sqrt{\frac{10.2/ps}{(\Delta M)_s}}
\qquad
\xi =
\frac{F_{B_s} \sqrt{B_{B_s}}}{F_{B_d} \sqrt{B_{B_d}}}
\end{equation}
where $\xi=1$ in the  SU(3)--flavour limit.
Note that $\mt$ and $|V_{cb}|$ dependences have been eliminated this way
 and that $\xi$ should in principle
contain much smaller theoretical
uncertainties than the hadronic matrix elements in $\Delta M_d$ and
$\Delta M_s$ separately.

The most recent values relevant for (\ref{107b}) are:
\begin{equation}\label{107c}
\Delta M_s > 10.2/ ps ~(95\%~{\rm  C.L.})
\qquad\quad
\xi=1.15\pm 0.05
\end{equation}
The first number is the improved lower bound from ALEPH \cite{Drell}.
The second number comes from quenched lattice calculations summarized
by Flynn in \cite{Flynn} and Bernard \cite{Bernard} here.
A similar result has been obtained using QCD sum rules \cite{NAR}.

The fate of the usefulness of the bound (\ref{107b}) depends clearly
on both $\Delta M_s$ and $\xi$.
For $\xi=1.2$
the lower bound on $\Delta M_s$ in (\ref{107c}) implies $R_t\le 1.20$
which, as we will see, has a moderate impact on the unitarity triangle
obtained on
the basis of the first four steps alone.

Finally, I would like to point out that whereas step 5 can give,
in contrast
to step 4, the value for $R_t$ free of the $\vcb$ uncertainty, it does
not provide at present a more accurate value of $\vtd$. The point is, that
having $R_t$, one determines $\vtd$ by means of the relation (\ref{106})
which, in contrast to (\ref{VT}), depends on $\vcb$.
In fact as we will see below, the inclusion
of step 5 has, with $\xi=1.2$, a visible impact on $R_t$ without
essentially any impact on the range of $\vtd$ obtained on the basis
of the first four steps alone.

\subsection{Numerical Results}
\subsubsection{Input Parameters}
 The input parameters needed to perform the
standard analysis using the first four steps alone
are given in table \ref{tab:inputparams}.
We list here the "present" errors based on what we have discussed
above, as well as the "future" errors. The latter are a mere guess,
but as we will see in the subsequent sections, these are the errors
one should aim at, in order that the standard analysis could be
competitive in the CKM determination with the cleanest rare decays and
the CP asymmetries in B-decays.

 $\mt$ in table 1
 refers
to the running current top quark mass normalized at $\mu=\mt$:
$\mtb(\mt)$. Its value given there corresponds to
$\mt^{Pole}=175\pm 6~GeV$ from CDF and D0.

\begin{table}[thb]
\caption[]{Collection of input parameters.\label{tab:inputparams}}
\vspace{0.4cm}
\begin{center}
\begin{tabular}{|c|c|c|c|}
\hline
{\bf Quantity} & {\bf Central} & {\bf Present} & {\bf Future} \\
\hline
$|V_{cb}|$ & 0.040 & $\pm 0.003$ & $\pm 0.001 $\\
$|V_{ub}/V_{cb}|$ & 0.080 & $\pm 0.020$ & $\pm 0.005 $ \\
$B_K$ & 0.75 & $\pm 0.15$ & $\pm 0.05$ \\
$\sqrt{B_d} F_{B_{d}}$ & $200\mev$ & $\pm 40\mev$ &$\pm 10\mev$ \\
$\mt$ & $167\gev$ & $\pm 6\gev$ & $\pm 3\gev $\\
$\Delta M_d$ & $0.464~\mbox{ps}^{-1}$ & $\pm 0.018~\mbox{ps}^{-1}$
& $\pm 0.006~\mbox{ps}^{-1}$\\
\hline
\end{tabular}
\end{center}
\end{table}
\noindent
{\bf Output of a Standard Analysis}

\noindent
The output of the standard analysis depends to some extent on the
error analysis. This should be always remembered in view of the fact
that different authors use different procedures. In order to illustrate
this  I show in tables \ref{TAB2} ("present") and \ref{TAB3} ("future")
the results for various quantities of interest
using two types of error analyses:

\begin{itemize}
\item
Scanning: Both the experimentally measured numbers and the theoretical input
parameters are scanned independently within the errors given in
table~\ref{tab:inputparams}.
\item
Gaussian: The experimentally measured numbers and the theoretical input
parameters are used with Gaussian errors.
\end{itemize}
Clearly the "scanning" method is a bit conservative. On the other
hand using Gaussian distributions for theoretical input parameters
can be certainly questioned.
I think that
at present the conservative "scanning" method should be preferred,
although one certainly would like to have a better method. An interesting
new method is  discussed by Schune in these proceedings \cite{FRENCH}.
The analysis discussed here has been done in collaboration with Matthias
Jamin and Markus Lautenbacher \cite{BJL96b}.

\begin{table}[th]
\caption[]{Present output of the Standard Analysis.
 $\lambda_t=V^*_{ts} V_{td}$.\label{TAB2}}
\vspace{0.4cm}
\begin{center}
\begin{tabular}{|c||c||c|}\hline
{\bf Quantity} & {\bf Scanning} & {\bf Gaussian} \\ \hline
$\mid V_{td}\mid/10^{-3}$ &$6.9 - 11.3$ &$ 8.6\pm 1.1$ \\ \hline
$\mid V_{ts}/V_{cb}\mid$ &$0.959 - 0.993$ &$0.976\pm 0.010$  \\ \hline
$\mid V_{td}/V_{ts}\mid$ &$0.16 - 0.31$ &$0.213\pm 0.034$  \\ \hline
$\sin(2\beta)$ &$0.36 - 0.80$ &$ 0.66\pm0.13 $ \\ \hline
$\sin(2\alpha)$ &$-0.76 - 1.0$ &$ 0.11\pm 0.55 $ \\ \hline
$\sin(\gamma)$ &$0.66 - 1.0 $ &$ 0.88\pm0.10 $ \\ \hline
$\IM \lambda_t/10^{-4}$ &$0.86 - 1.71 $ &$ 1.29\pm 0.22 $ \\ \hline
\end{tabular}
\end{center}
\end{table}

\begin{table}[hb]
\caption[]{Future output of the Standard Analysis.
 $\lambda_t=V^*_{ts} V_{td}$.\label{TAB3}}
\vspace{0.4cm}
\begin{center}
\begin{tabular}{|c||c||c|}\hline
{\bf Quantity} & {\bf Scanning} & {\bf Gaussian} \\ \hline
$\mid V_{td}\mid/10^{-3}$ &$8.1 - 9.2$ &$ 8.6\pm 0.3$ \\ \hline
$\mid V_{ts}/V_{cb}\mid$ &$0.969 - 0.983$ &$0.976\pm 0.004$  \\ \hline
$\mid V_{td}/V_{ts}\mid$ &$0.20 - 0.24$ &$0.215\pm 0.010$  \\ \hline
$\sin(2\beta)$ &$0.61 - 0.70$ &$ 0.67\pm0.03 $ \\ \hline
$\sin(2\alpha)$ &$-0.11 - 0.66.0$ &$ 0.21\pm 0.21 $ \\ \hline
$\sin(\gamma)$ &$0.90 - 1.0 $ &$ 0.96\pm0.03 $ \\ \hline
$\IM \lambda_t/10^{-4}$ &$1.21 - 1.41 $ &$ 1.29\pm 0.06 $ \\ \hline
\end{tabular}
\end{center}
\end{table}

\begin{figure}[hbt]
\vspace{0.10in}
\centerline{
\epsfysize=3.4in
\rotate[r]{\epsffile{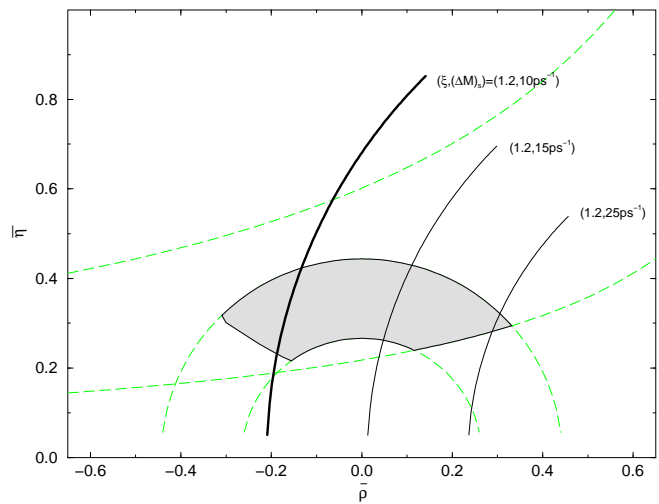}}
}
\vspace{0.08in}
\caption[]{
Unitarity Triangle 1997.
\label{fig:utdata}}
\end{figure}

\begin{figure}[hbt]
\vspace{0.10in}
\centerline{
\epsfysize=3.6in
\rotate[r]{\epsffile{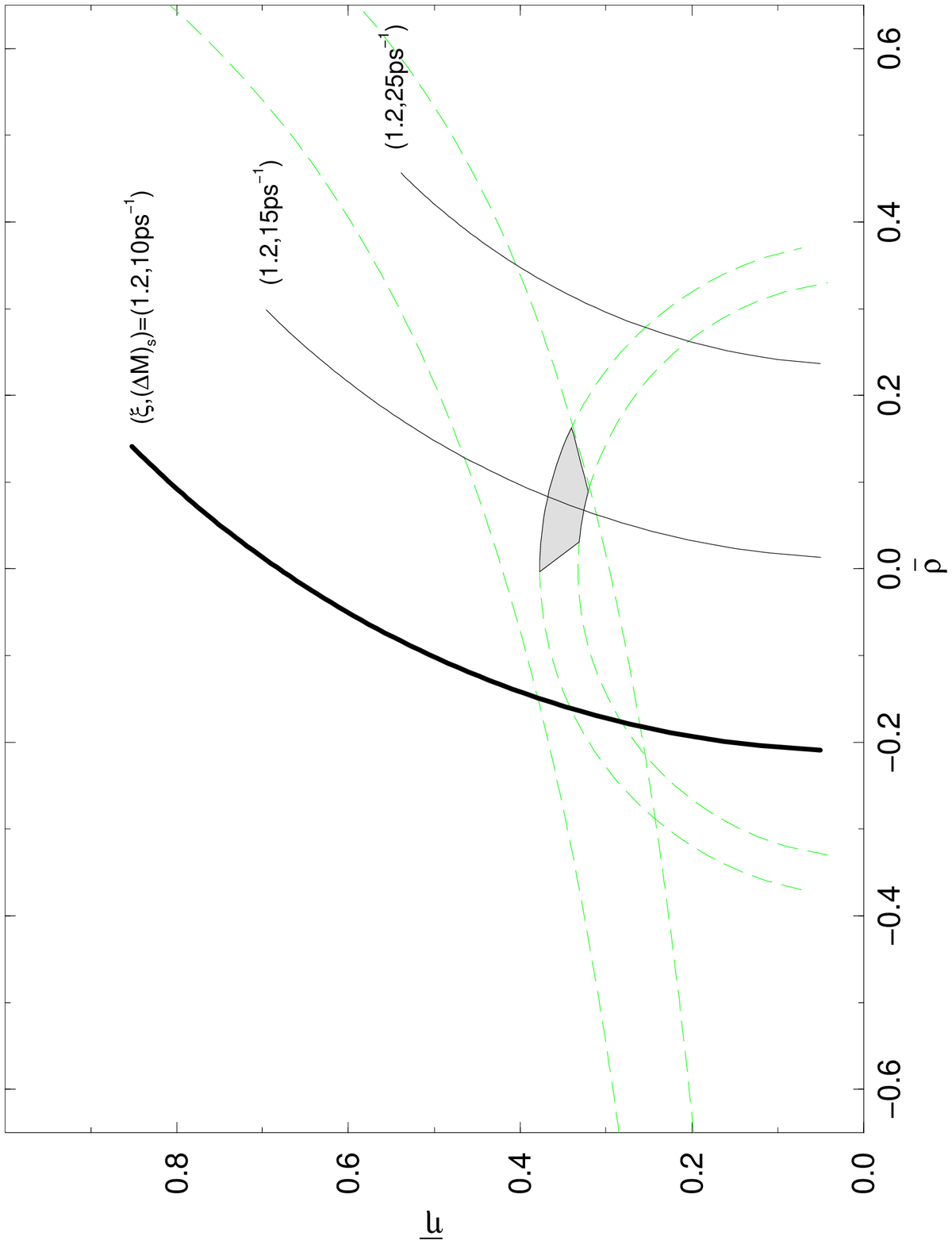}}
}
\vspace{0.08in}
\caption[]{
Unitarity Triangle 2007.
\label{fig:utdataf}}
\end{figure}

In figs. \ref{fig:utdata} and  \ref{fig:utdataf}  we show the ranges
 for the upper
corner A of the UT in the case of the "present" input and "future" input
respectively. The circles correspond to $R_t^{max}$ from
(\ref{107b})
using $\xi=1.20$ and $(\Delta M)_s=10/ps,~15/ps$ and $25/ps$, respectively.
The present bound (\ref{107c}) is represented by the first
of these circles.
This bound has not
been used in
obtaining the results in tables \ref{TAB2} and \ref{TAB3}.
Its impact will be analysed separately
below.
The circles from $B^0_d-\bar B^0_d$ mixing are not shown explicitly
for reasons to be explained below. The impact of $\Delta M_d$ can however
be easily seen by comparing the shaded area with the area one would find
by using the lower $\varepsilon$-hyperbola and the $R_b$-circles alone.

The allowed region has a typical "banana" shape which can be found
in many other analyses \cite{BLO,ciuchini:95,HNb,ALUT,FRENCH}. The size of
the banana and its position depends on the assumed input
parameters and on the error analysis which varies from paper
to paper. The results in figs. \ref{fig:utdata} and  \ref{fig:utdataf}
correspond to a simple independent
scanning of all parameters within one standard deviation.

As seen in fig. \ref{fig:utdata} our present knowledge of
the unitarity triangle is still rather poor. Fig. \ref{fig:utdataf}
demonstrates very clearly that this situation may change dramatically
in the future provided the errors in the input parameters will be decreased
as anticipated in our "future" scenario.

Comparing the results for $\vtd$ given in table \ref{TAB2}
with the ones obtained on the basis of unitarity alone (\ref{uni1})
we observe that
the inclusion of the constraints from $\varepsilon$ and $\Delta M_d$
had a considerable impact on the allowed range for this CKM matrix
element. This impact will be amplified in the future as seen in
table \ref{TAB3}. An inspection shows that with our input parameters
the lower bound on $\vtd$ is governed by  $\varepsilon_K$, whereas
the upper bound by $\Delta M_d$.

Next we observe that whereas the angle $\beta$ is rather
constrained, the uncertainties in
$\alpha$ and $\gamma$ are  huge:
\be\label{ap}
35^\circ\le \alpha \le 115^\circ
\quad
11^\circ\le \beta \le 27^\circ
\quad
41^\circ\le \gamma \le 134^\circ
\ee
The situation will improve when the "future" scenario
will be realized:
\be\label{af}
70^\circ\le \alpha \le 93^\circ
\quad
19^\circ\le \beta \le 22^\circ
\quad
65^\circ\le \gamma \le 90^\circ
\ee

\noindent
{\bf Impact of $\Delta M_s$}

\noindent
The impact of the present lower bound on $\Delta M_s$ is still
very small except for the upper limits for $\vtd/\vts$ and $\gamma$
which are lowered in the "scanning'' version to $0.27$ and $129^\circ$
respectively.
\subsubsection{Correlation between $\varepsilon_K$ and $\Delta M_d$}
Now, why did we omitt the explicit circles from $B^0_d-\bar B^0_d$ mixing
in the plots of unitarity triangles above ? I have to answer this
question because some of my colleagues suspected that a plot similar
to the one in fig. \ref{fig:utdata} and shown already at the Rochester
conference in Warsaw was wrong. At first  one would expect that the
left border of the allowed area coming from $B^0_d-\bar B^0_d$ mixing
should have a shape similar to the circles coming from
$\Delta M_d/\Delta M_s$ and shown in the figures above. This expectation
is correct at fixed values of $m_t$ and $\vcb$. Yet once these
two parameters are varied in the allowed ranges, this is no longer
true. In fact one can easily convince oneself that the uncertainties
coming from $\mt$ and $\vcb$ in the analyses of $\varepsilon_K$ and
$\Delta M_d$ cannot be represented simultaneously in the
$(\bar\varrho,\bar\eta)$ plane in terms of nice hyperbolas
and nice circles. This is simply related to the correlation between
$\varepsilon_K$ and $\Delta M_d$ due to $m_t$ and $\vcb$. Neglecting
this correlation one finds for instance that the most negative value of
$\bar\varrho$ corresponds to the maximal values of $(m_t,\vcb)$ in the
case of $\varepsilon_K$ and to the minimal values of $(m_t,\vcb)$ in the
case of $B^0_d-\bar B^0_d$ which is of course inconsistent. In
figs. \ref{fig:utdata} and  \ref{fig:utdataf} we have decided
to show the $\varepsilon_K$-hyperbolas. Consequently the impact
of $B^0_d-\bar B^0_d$ mixing had to be found numerically and as
seen it is not described by a circle. Since $m_t$ is already
very well known, this discussion mainly applies to the $\vcb$ dependence.
Finally it should be stressed that similar correlations have to
be taken into account in the future when various rare decays discussed in
the next section will enter the game of the determination of the
unitarity triangle. Needless to say, the radius $R^{max}_t$
determined through
(\ref{107b}) and shown in the UT plots, being independent of
$(\mt,\vcb)$, is not subject to the correlation in question.

\section{CKM from Rare  Decays}
\subsection{Preliminary Remarks}
Let us change the gears and move to rare K- and B-decays. Doing this
we actually move into the future, a very interesting one as we will
see.
Not all rare decays are suitable, at least at present,
 for precise determinations of the
CKM parameters. Several of them suffer from large hadronic uncertainties
and only dramatic progress in non-perturbative techniques could
change them to useful means for CKM determinations. On the other hand
there is a small number of decays with small or even negligible
theoretical uncertainties. These decays are particularly suitable
for the determination of the CKM matrix. Here the main difficulty
are
very small branching ratios and the fate of the usefulness of these
decays after the completion of NLO-QCD corrections \cite{BBL}
lies mainly in the hands of experimentalists.

We will begin our discussion with decays which have large
theoretical uncertainties. Subsequently we will gradually move towards
cleaner decays, ending our discussion with the {\it gold plated decay}
$K_L\to \pi^0\nu\bar\nu$ and its {\it silver plated sister}
$K^+\to\pi^+\nu\bar\nu$.

\subsection{$\epe$, $K_L\to \mu\bar\mu$, $B \to K^*(\varrho)\gamma$ and
 $K_L\to\pi^0 \lowercase{e}^+\lowercase{e}^-$}

$\epe$ suffers from large
uncertainties in the relevant hadronic matrix elements of
QCD-penguin and electroweak penguin operators \cite{BF97}.
The situation is further complicated by the strong cancellations
between these contributions. This precludes a useful extraction of
$\IM\lambda_t$ from future data unless some dramatic progress
in non-perturbative calculations will take place.
In spite of this a measurement of $\varepsilon'/\varepsilon$ at
the $10^{-4}$ level remains as one of the important targets of
contemporary particle  physics. A non-vanishing value of this ratio
would give the first signal for the direct CP violation ruling out
the superweak models \cite{WO1}.
The clarification of this important issue should come soon from
FNAL, CERN and at a later stage from
DA$\Phi$NE.

The branching ratio for the decay $K_L \to \mu\bar\mu$ has been
already measured \cite{PRINZ}.
Unfortunately,
it is dominated by long distance contributions. The extraction of the
well calculable short distance
contribution ( sensitive to $\bar\varrho$) from these data
remains a very important challenge.
A recent discussion of this issue and the relevant references can be found
in \cite{DIP}.

The exclusive decays $B \to K^*(\varrho)\gamma$ are known experimentally
from CLEO measurements. Taken together they provide an upper bound
on $\vtd$, which is, however, very weak. The strong model dependence in
the relevant formfactors precludes a useful determination of $\vtd$
from these decays.
This is at least my opinion. There are others who are more optimistic
in this respect. Clearly a considerable progress in lattice calculations
could improve this situation.

The decay $K_L\to \pi^0 e^+e^-$ receives three contributions:
CP conserving, {\it indirectly} CP violating and {\it directly}
CP violating. The directly CP violating part can be calculated
very reliably:
$Br(K_L\to\pi^0 e^+ e^-)_{dir}=(4.5 \pm 2.6)\cdot 10^{-12}$,
where the error comes dominantly from the uncertainties in the CKM
parameters. Extracting this part from future data would give a clean
determination of  $\IM\lambda_t$.
Unfortunately the other two components, although likely smaller, are
not fully negligible.
A better assessment of the importance of the
indirect CP violation in $K_L\to\pi^0e^+e^-$ will become possible after
the measurement of $Br(K_S\to\pi^0e^+e^-)$
at DA${\Phi}$NE.
On the other hand the prospects of the accurate estimate of the
CP conserving part are less clear.
The present experimental bound:
$Br(K_L\to\pi^0 e^+ e^-) < 4.3 \cdot 10^{-9}$ \cite{harris}
should be considerably improved in the coming years at FNAL.
More details on this interesting decay can
be found in  \cite{PICH97,BF97}.

\subsection{$B \to X_{d,s}\gamma$ and $B\to X_{d,s} l^+l^-$}
These decays are covered by Ali \cite{ALI97} and Hewett in these
proceedings.
Let me still
make a few comments on them.

The decay $B \to X_{s}\gamma$ has been a subject of intensive
efforts by experimentalists and theorists during the last 5-10
years.
Experimentally its branching ratio is found to be \cite{CLEO2}:
\begin{equation}\label{EXP}
Br(B \to X_s\gamma) = (2.32 \pm 0.57 \pm 0.35) \times 10^{-4}\,,
\end{equation}
and a very preliminary result from ALEPH reads
$(3.38 \pm 0.74 \pm 0.85) \times 10^{-4}$.
Here the first errors are statistical and the second  systematic.
On the other hand the Standard Model NLO analyses
\cite{BKP1} give
\be\label{sfin}
Br(B{\to}X_s \gamma) =(3.48 \pm 0.13~({\rm scale})~\pm 0.28~({\rm par}))
  \times 10^{-4}
= (3.48 \pm 0.31)  \times 10^{-4}
\ee
where this particular estimate comes from the paper by Kwiatkowski,
Pott and myself.
We show separetely scale and parametric uncertainties.

The error in the theoretical estimate is
dominated by paramatric uncertainties of which the quark masses and
the used branching ratio $Br(B \to X_c e \nu_e)$ are most important ones.
I believe that these parametric uncertainties will be reduced in the
future by at least a factor of two, so that a prediction for
$Br(B{\to}X_s \gamma)$ with an uncertainty of $5-10\%$ will be
available within the next, say, five years. Provided the experimental
 branching ratio can be measured with a similar accuracy this would
allow a $5\%$ measurement of the ratio $|V_{ts}/V_{cb}|$. Even if such
a measurement would not improve our knowledge of this ratio within the
Standard Model, its considerable departure from 0.975 would signal
the physics beyond the Standard Model. To achieve the accuracy of
$5\%$ in the experimental branching ratio is of course a very
difficult but not a fully unrealistic task.
{}From the existing data Ali extracts $\vts=0.033\pm0.007$.

The issue of $B\to X_d\gamma$, sensitive to $\vtd$, is a different story.
It is much harder to measure than $B\to X_s\gamma$ and the analysis
is not as simple as in the latter case because the CKM non-leading
contributions, which are negligible in $B\to X_s\gamma$, have
to be considered now \cite{ALI97}. Even if there are different opinions about
this, many would
agree with me that this decay cannot compete with the measurement
of $\vtd$ through $K^+\to\pi^+\nu\bar\nu$ or even through
$B^0_d-\bar B_d$ mixing (provided $\sqrt{B_d}F_{B_d}$ will be improved).
Yet the efforts should be
made to measure it because as in the case of $B\to X_s\gamma$
its branching ratio is sensitive to the physics beyond the Standard
Model.

Similar comments apply to $B \to X_{s,d} l^-l^+$ decays except
that here one is faced with an additional  obstacle: the long distance
contributions, due mainly to $J/\psi$ and $\psi'$ resonances.
An interesting new study of this issue can be found in \cite{BBSNU}.
These difficulties can be avoided to some extent
by studing special
regions of the invariant dilepton mass spectrum or asymmetries of
various sorts. These issues are discussed by Ali and Hewett in these
proceedings.
The extraction of the CKM parameters from these decays has been
discussed recently in \cite{KMS97}.

\subsection{ $B_{d,s}\to\mu\bar\mu$ and $B\to X_{d,s}\nu\bar\nu$}

$B_{d,s}\to\mu\bar\mu$ and $B\to X_{d,s}\nu\bar\nu$ are the theoretically
cleanest decays in the field of rare B-decays.
They are dominated by the $Z^0$-penguin and box diagrams
involving top quark exchanges.
After the calculation of NLO-QCD corrections \cite{BB13}, the scale
uncertainties in these decays are negligible.
The same applies to long distance contributions \cite{BBSNU}.
One has then:
\begin{equation}
Br(B_s\to \mu\bar\mu)=
3.5\cdot 10^{-9}\left[\frac{F_{B_s}}{210~MeV}\right]^2
\left[\frac{\mtb(m_t)}{170~GeV} \right]^{3.12}
\left[\frac{\mid V_{ts}\mid}{0.040} \right]^2
\left[\frac{\tau_{B_s}}{1.6 ps} \right]
\end{equation}
and
\begin{equation}
Br(B\to X_s\nu\bar\nu)=
3.7\cdot 10^{-5}
\left[\frac{|V_{ts}|}{|V_{cb}|} \right]^2
\left[\frac{\mtb(m_t)}{170~GeV} \right]^{2.3}
\left[\frac{0.54}{f(z)}\right]
\left[\frac{Br(B \to X_c e \bar{\nu}_e)}{0.104} \right]
 \end{equation}
where $f(z)$ is a phase space factor with $z=m_c/m_b$.

The short
distance character of these decays allows
a clean determination of $\vtd/\vts$
by measuring the ratios $Br(B_d\to \mu\bar\mu)/Br(B_s\to \mu\bar\mu)$
and $Br(B\to X_d \nu\bar\nu)/Br(B\to X_s\nu\bar\nu)$. In particular
the latter determination is very clean as the uncertainties related to
$Br(B \to X_c e \bar{\nu}_e)$ and $f(z)$ cancel in the ratio.
The corresponding determination using $B_{d,s}\to \mu\bar\mu$
suffers from the uncertainty in the ratio $F_{B_d}/F_{B_s}$ which
however should be removed to a large extend by future lattice
calculations.

Experimentally there exists an indirect upper bound (90\% C.L.):
$Br(B\to X_s \nu\bar\nu) < 7.7\cdot 10^{-4}$
obtained by ALEPH in 1996.
Yet it is fair to say that the actual measurements of this branching ratio
and in particular of $Br(B\to X_d \nu\bar\nu)$ are extremely difficult.
The measurement of $B_{s,d}\to \mu\bar\mu$ is
first of all difficult because of the expected tiny branching ratio.
Still one should hope that our experimental colleagues and those who
provide financial support will surprise us and some useful branching
ratios for these clean decays will be available before 2011.

\subsection{$K_L\to\pi^0\nu\bar\nu$ and $K^+\to\pi^+\nu\bar\nu$}
$K_L\to\pi^0\nu\bar\nu$ and $K^+\to\pi^+\nu\bar\nu$ are the theoretically
cleanest decays in the field of rare K-decays.
$K_L\to\pi^0\nu\bar\nu$ is
dominated by Z-penguins and box diagrams
involving the top quark.  $K^+\to\pi^+\nu\bar\nu$ receives
additional sizable contributions from internal charm exchanges.
The great virtue of $K_L\to\pi^0\nu\bar\nu$ is that it proceeds
almost exclusively through direct CP violation \cite{Littenberg}
and as such is the
cleanest decay to measure this important phenomenon. It also offers
a clean determination of the Wolfenstein parameter $\eta$ and in particular
as we will stress below offers the cleanest measurement
of $\IM\lambda_t= \IM V^*_{ts} V_{td}$.
$K^+\to\pi^+\nu\bar\nu$ is CP conserving and offers a clean
determination of $|V_{td}|$. Due to the presence of the charm
contribution  it has a non-negligible scale uncertainty and
an uncertainty in $m_c$ both  absent in $K_L\to\pi^0\nu\bar\nu$.

After the NLO-QCD calculations \cite{BB13},
the remaining scale uncertainty
in the extraction of $\vtd$ from
$Br(K^+ \rightarrow \pi^+ \nu \bar{\nu})$ amounts to
$\pm 4\%$.
The remaining scale uncertainty
in the determination of $\eta$
or $\IM\lambda_t$ from $Br(K_L\to\pi^0\nu\bar\nu)$
is below $\pm 1\%$ which is safely negligible.
Since the relevant hadronic matrix
elements of the weak currents entering $K\to \pi\nu\bar\nu$
can be related using isospin symmetry to the leading
decay $K^+ \rightarrow \pi^0 e^+ \nu$, the resulting theoretical
expressions for Br( $K_L\to\pi^0\nu\bar\nu$) and Br($K^+\to\pi^+\nu\bar\nu$)
do not contain any hadronic uncertainties.
The isospin braking corrections have been calculated in
\cite{MP}.
The long distance contributions to
$K^+ \rightarrow \pi^+ \nu \bar{\nu}$ have been
considered in \cite{RS} and found to be
safely neglegible.
The long distance contributions to $K_L\to\pi^0\nu\bar\nu$ are negligible
as well. Finally
the two-loop
($\ord (G_F^2\mt^4)$ at the amplitude level) electroweak corrections
to $K_L\to\pi^0\nu\bar\nu$
have been computed \cite{BB97} and found
to be about
$2\%$ which is  well below the experimental sensitivity in the
forseeable future.

The explicit expressions for $Br(K^+ \rightarrow \pi^+ \nu \bar{\nu})$
and $Br(K_L\to\pi^0\nu\bar\nu)$ can be found in \cite{BBL,BF97}.
In particular:
\begin{equation}\label{bklpn}
Br(K_L\to\pi^0\nu\bar\nu)=
3.0\cdot 10^{-11}\left [ \frac{\eta}{0.39}\right ]^2
\left [\frac{\mtb(\mt)}{170~GeV} \right ]^{2.3}
\left [\frac{\mid V_{cb}\mid}{0.040} \right ]^4.
\end{equation}
Scanning the "present" input parameters of table 1 one finds
\cite{BJL96b}:
\begin{equation}
Br(K^+ \rightarrow \pi^+ \nu \bar{\nu})=
(9.1\pm 3.8)\cdot 10^{-11}\quad,\quad
Br(K_L\to\pi^0\nu\bar\nu)=(2.8\pm 1.7)\cdot 10^{-11}
\end{equation}
where the errors are dominated by the present uncertainties in
$\vcb$, $\vtd$ and $\eta$. These errors are decreased by roughly
a factor of three if the "future" input parameters are used.

One of the high-lights of August 97 was the observation by BNL787
\cite{Adler97}
of one event consistent with the signature expected for this decay.
The branching ratio:
\be\label{kp97}
Br(K^+ \rightarrow \pi^+ \nu \bar{\nu})=
(4.2+9.7-3.5)\cdot 10^{-10}
\end{equation}
has the central value  by a factor of 4 above the Standard Model
expectation but in view of large errors the result is compatible with the
Standard Model. This new result implies that $\vtd$ lies in the range
$0.006<\vtd< 0.06 $ which is substantially larger than the range
from the standard analysis of section 3. The analysis of additional
data on $K^+\to \pi^+\nu\bar\nu$ present on tape at BNL787 should narrow
this range in the near future considerably.
It is also hoped that FNAL will contribute to the measurement of
this branching ratio \cite{Cooper}.

The most recent upper bound on $Br(K_L\to \pi^0\nu\bar\nu)$ from
FNAL-E799 is $1.8 \cdot 10^{-6}$.
This experiment expects to reach
the accuracy ${\cal O}(10^{-8})$ and a very interesting new proposal
 AGS2000 \cite{AGS2000}
expects to reach the single event sensitivity $2\cdot 10^{-12}$
allowing a $10\%$ measurement of the expected branching ratio.
It is hoped that also JNAF, FNAL and KEK will make efforts to measure
this gold-plated decay.
\subsection{$\vtd$, $\IM\lambda_t$, $\sin 2\beta$ and UT from
$K \to \pi\nu\bar\nu$}

\subsubsection{$\vtd$ from $K^+\to\pi^+\nu\bar\nu$}
Once $Br(K^+\to\pi^+\nu\bar\nu)\equiv Br(K^+)$ is measured, $\vtd$ can be
extracted subject to various uncertainties:
\be\label{vtda}
\frac{\sigma(\vtd)}{\vtd}=\pm 0.04_{scale}\pm \frac{\sigma(\vcb)}{\vcb}
\pm 0.7 \frac{\sigma(\bar\mc)}{\bar\mc}
\pm 0.65 \frac{\sigma( Br(K^+))}{Br(K^+)}
\ee
Taking $\sigma(\vcb)=0.002$, $\sigma(\bar\mc)=100\mev$ and
$\sigma( Br(K^+))=10\%$ and adding the errors in quadrature we find that
$\vtd$ can be determined with an accuracy of $\pm 10\%$ in the present
example. This number
is increased to $\pm 11\%$ once the uncertainties due to $\mt$,
$\alpha_s$ and $|V_{ub}|/\vcb$ are taken into account. Clearly this
measurement can be improved although a determination of $\vtd$ with
an accuracy better than $\pm 5\%$ seems rather unrealistic.

\subsubsection{$\IM \lambda_t$ and $\vcb$ from $K_L\to\pi^0\nu\bar\nu$}
The basic formulae here are \cite{BB96,AJB94}:
\begin{equation}\label{imlta}
\IM\lambda_t=1.36\cdot 10^{-4}
\left[\frac{170\gev}{\mtb(\mt)}\right]^{1.15}
\left[\frac{Br(\klpn)}{3\cdot 10^{-11}}\right]^{1/2}\,.
\end{equation}
and
\begin{equation}\label{vcbklpn}
|V_{cb}|=40.0\cdot 10^{-3} \sqrt{\frac{0.39}{\eta}}
\left[\frac{170\gev}{\mtb(\mt)}\right]^{0.575}
\left[\frac{Br(\klpn)}{3\cdot 10^{-11}}\right]^{1/4}\,.
\end{equation}

(\ref{imlta}) offers
 the cleanest method to measure $\IM\lambda_t$;
even better than the CP asymmetries
in $B$ decays discussed in the following section.
On the other hand (\ref{vcbklpn})
determines $\vcb$ without any hadronic
uncertainties \cite{AJB94}.
With $\eta$
determined one day from CP asymmetries in B-decays
and $\mt$ measured very precisely at LHC and NLC,
a measurement of $Br(\klpn)$ with an accuracy of $10\%$
would determine $\vcb$ with an error of $\pm 0.001$.
A comparision of
this determination of $|V_{cb}|$ with the usual one in tree level
B-decays would offer an excellent test of the Standard Model
and in the case of discrepancy would signal physics beyond
it.

\begin{figure}[hbt]
\vspace{0.10in}
\centerline{
\epsfysize=1.7in
\epsffile{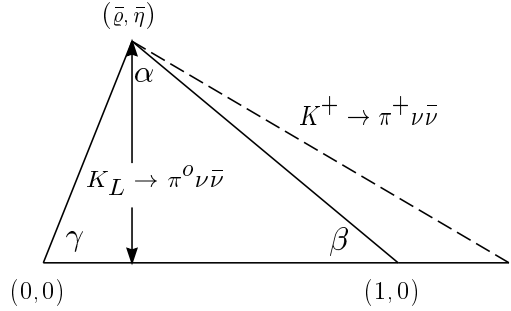}
}
\vspace{0.02in}
\caption{Unitarity triangle from $K\to\pi\nu\bar\nu$.}\label{fig:KPKL}
\end{figure}
\noindent
{\bf Unitarity Triangle and $\sin 2\beta$ }

\noindent
The measurements of $Br(\kpn)$ and $Br(\klpn)$ can determine the
unitarity triangle completely, (see fig. \ref{fig:KPKL}),
provided $\mt$ and $V_{cb}$ are known \cite{BB4}.
Of particular interest is the determination of $\sin 2\beta$
\cite{BB4} which is independent of $\vcb$ and $\mt$:
\begin{equation}\label{sin}
\sin 2\beta=\frac{2 r_s}{1+r^2_s}
\qquad
r_s={\sqrt{B_1-B_2}-P_c\over\sqrt{B_2}}\,.
\end{equation}
where
\begin{equation}\label{b1b2}
B_1={Br(\kpn)\over 4.11\cdot 10^{-11}}\qquad
B_2={Br(\klpn)\over 1.80\cdot 10^{-10}}\,.
\end{equation}
and $P_c=0.40\pm0.06$ represents the charm contribution to
$K^+\to\pi^+\nu\bar\nu$.

An extensive numerical analysis of the unitarity triangle from
$K \to \pi\nu\bar\nu$ can be found in
 \cite{BB4,BB96}. Here we only briefly compare this determination with
the one by means of the standard analysis of the unitarity triangle.
We  assume
that $Br(\kpn)$ and
$Br(\klpn)$ are known to within $\pm 10\%$, $\sigma (\mc)= \pm 50\mev$ and
$\sigma (\mt)= \pm 3\gev$. Then for two choices of the uncertainty in
$\vcb$
one finds the results given in the second and the third column of table
\ref{tabkb1}. In the fourth and fifth column the corresponding results
of the standard analysis of the unitarity triangle are shown.
We observe that very respectable determinations of all considered
quantities except for
$\bar\varrho$ can be obtained from $K \to \pi\nu\bar\nu$.
However even in this case there is a considerable progress
when compared with the present analysis of the unitarity triangle.
Of particular interest are the accurate determinations of $\sin 2\beta$
and $\IM\lambda_t$.
The comparision of this determination of $\sin 2\beta$ with the
one through the CP asymmetry in $B_d\to \psi K_{\rm S}$
is particularly suited for tests of CP violation in the
Standard Model and offers a powerful tool to probe the physics
beyond it.
The discussion of $K_L\to\pi^0\nu\bar\nu$ beyond the Standard Model
can be found in \cite{KLBSM}

\begin{table}
\caption[]{Illustrative example of the determination of CKM
parameters from $K\to\pi\nu\bar\nu$ and from the standard
analysis of the unitarity triangle.
\label{tabkb1}}
\vspace{0.4cm}
\begin{center}
\begin{tabular}{|c||c|c||c|c|}\hline
&$\sigma(|V_{cb}|)=\pm 0.002$ & $\sigma(|V_{cb}|)=\pm 0.001$
& {\rm Present} & {\rm Future}
\\
\hline
\hline
$\sigma(|V_{td}|) $& $\pm 10\% $ & $ \pm 9\% $
& $\pm 24\%$ & $\pm 7\%$\\
\hline
$\sigma(\bar\varrho) $ & $\pm 0.16$ &$\pm 0.12$
& $\pm 0.32$  & $\pm 0.08$\\
\hline
$\sigma(\bar\eta)$ & $\pm 0.04$&$\pm 0.03$
&$\pm 0.12 $ & $\pm 0.03 $\\
\hline
$\sigma(\sin 2\beta)$ & $\pm 0.05$&$\pm 0.05$
& $\pm 0.22 $ & $\pm 0.05$\\
\hline
$\sigma({\rm Im}\lambda_t)$&$\pm 5\%$ &$\pm 5\%$
& $\pm 33\%$ & $\pm 8\%$\\
\hline
\end{tabular}
\end{center}
\end{table}

\section{$\alpha$, $\beta$ and $\gamma$ from two-body B-Decays}
\subsection{CP-Asymmetries in B-Decays}
CP violation in B decays is certainly one of the most important
targets of B factories and of dedicated B experiments at hadron
facilities. It is well known that CP violating effects are expected
to occur in a large number of channels at a level attainable at
forthcoming experiments. Moreover there exist channels which
offer the determination of CKM phases essentially without any hadronic
uncertainties. Since CP violation in B decays has been already
reviewed in two special talks by Fleischer and Sanda at this symposium
and since extensive reviews can be found in the literature
\cite{NQ,RFD,BF97}, let me concentrate only on the most important points.

The classic determination of $\alpha$ by means of the
time dependent CP  asymmetry in the decay
$B_d^0 \rightarrow \pi^+ \pi^-$
is affected by the "QCD penguin pollution" which has to be
taken care of in order to extract $\alpha$.
The recent CLEO results for penguin dominated decays indicate that
this pollution could be substantial.
The most popular strategy to deal with this "penguin problem''
is the isospin analysis of Gronau and London \cite{CPASYM}. It
requires however the measurement of $Br(B^0\to \pi^0\pi^0)$ which is
expected to be below $10^{-6}$: a very difficult experimental task.
For this reason several, rather involved, strategies \cite{SNYD}
have been proposed which
avoid the use of $B_d \to \pi^0\pi^0$ in conjunction with
$a_{CP}(\pi^+\pi^-,t)$. They are reviewed in \cite{BF97}.
 It is to be seen which of these methods
will eventually allow us to measure $\alpha$ with a respectable precision.
It is however clear that the determination of this angle is a real
challenge for both theorists and experimentalists.

The CP-asymmetry in the decay $B_d \rightarrow \psi K_S$ allows
 in the Standard Model
a direct measurement of the angle $\beta$ in the unitarity triangle
without any theoretical uncertainties \cite {BSANDA}.
Of considerable interest \cite{RFD,PHI} is also the pure penguin decay
$B_d \rightarrow \phi K_S$, which is expected to be sensitive
to physics beyond the Standard Model. Comparision of $\beta$
extracted from $B_d \rightarrow \phi K_S$ with the one from
$B_d \rightarrow \psi K_S$ should be important in this
respect.

The two theoretically cleanest methods for the determination of $\gamma$
are: i) the full time dependent analysis of
$B_s\to D^+_s K^{-}$ and $\bar B_s\to D^-_s K^{+}$  \cite{adk}
and ii) the known triangle construction due to Gronau and Wyler \cite{Wyler}
which uses six decay rates $B^{\pm}\to D^0_{CP} K^{\pm}$,
$B^+ \to D^0 K^+,~ \bar D^0 K^+$ and  $B^- \to D^0 K^-,~ \bar D^0 K^-$.
Both methods are  unaffected by penguin contributions.
The first method is experimentally very
challenging because of the
expected large $B^0_s-\bar B^0_s$ mixing. The second method is problematic
because of the small
branching ratios of the colour supressed channels $B^{\pm}\to D^0 K^{\pm}$
giving a rather squashed triangle and thereby
making
the extraction of $\gamma$ difficult. Variants of the latter method
which could be more promising have been proposed in \cite{DUN2,V97}.
It appears that these methods will give useful results at later stages
of CP-B investigations. In particular the first method will be feasible
only at LHC-B.

For the time being the most promising for the extraction of $\gamma$
appears to be the method of Fleischer \cite{F96} and of
Fleischer and Mannel \cite{FM97} which
uses the rates for $B^{\pm}\to\pi^{\pm}K$ and $B_d\to\pi^{\mp}K^{\pm}$.
This method implies the bound:
\be
\sin^2\gamma \le
\frac{Br(B_d\to\pi^{\mp}K^{\pm})}{Br(B^{\pm}\to\pi^{\pm}K)}\equiv R
\ee
The Fleischer-Mannel bound is of particular interest because the most
recent CLEO data give $R=0.65\pm 0.40$ \cite{CLEO97}.
The firm conclusion cannot
be reached at present because of substantial experimental error.
However if improved data will continue to give $R<1$, the FR bound
would exclude the region around $\bar\varrho=0$ in the
$(\bar\varrho,\bar\eta)$ space  putting
the "$\gamma=90^\circ$ club" \cite{BjSt} into serious difficulties. It
should be stressed that excluding the region around $\bar\varrho=0$
would have a profound impact on the unitarity triangle dividing the
allowed region for its apex into well separated regions with
$\bar\varrho<0$ and $\bar\varrho>0$. The former could then probably
be eliminated by improving the lower bound on $\Delta M_s$ leaving
only a small allowed area with $\bar\varrho>0$. The crucial question
then is, whether R is indeed smaller than unity ? Hopefully CLEO
will answer this question in the coming years. More details on
the implications of the FR bound can be found in \cite{FM97,F97,GNF}.

\subsection{UT from CP-B and $K\to\pi\nu\bar\nu$ }
In what follows let us assume that $\alpha$, $\beta$ and $\gamma$
have been measured to certain accuracy and let us ask what such
measurement will imply for $\vtd$, $\bar\varrho$, $\bar\eta$
and $\IM\lambda_t$. To this end let us assume \cite{BB96}
\begin{equation}\label{sin2a2bI}
\sin 2\alpha=0.40\pm 0.10 \qquad \sin 2\beta=0.70\pm 0.06
\qquad ({\rm scenario\ I})
\end{equation}
\begin{equation}\label{sin2a2bII}
\sin 2\alpha=0.40\pm 0.04 \qquad \sin 2\beta=0.70\pm 0.02
\qquad ({\rm scenario\ II})
\end{equation}
Scenario I corresponds to the accuracy being aimed for at $B$-factories,
 HERA-B and FNAL prior to the LHC era.
An improved precision can be anticipated from
LHC experiments, which we illustrate with our choice of scenario II.
The assumed accuracy on $\alpha$ in the scenarion II is probably
unrealistic in view of the comments made above but let us try it anyway.
In general the calculation of $\bar\varrho$ and $\bar\eta$ from
$\sin 2\alpha$ and $\sin 2\beta$ involves discrete ambiguities
\cite{NIRW}. We will
assume that they can be removed by looking at other decays.

\begin{table}
\caption[]{Illustrative example of the determination of CKM
parameters from $K\to\pi\nu\bar\nu$ and B-decays.
\label{tabkb}}
\vspace{0.4cm}
\begin{center}
\begin{tabular}{|c||c||c|c|}\hline
&$K\to\pi\nu\bar\nu$
& {\rm Scenario I} & {\rm Scenario II}
\\
\hline
\hline
$\sigma(|V_{td}|) $& $\pm 10\% (9\% )$
& $\pm 5.5\% (3.5\%)$ & $\pm 5.0\% (2.5\%)$\\
\hline
$\sigma(\bar\varrho) $ & $\pm 0.16 (0.12)$
& $\pm 0.03$  & $\pm 0.01$\\
\hline
$\sigma(\bar\eta)$ & $\pm 0.04(0.03)$
&$\pm 0.04 $ & $\pm 0.01 $\\
\hline
$\sigma(\sin 2\beta)$ & $\pm 0.05$
& $\pm 0.06 $ & $\pm 0.02$\\
\hline
$\sigma({\rm Im}\lambda_t)$&$\pm 5\%$
& $\pm 14\%(11\%)$ & $\pm 10\%(6\%)$\\
\hline
\end{tabular}
\end{center}
\end{table}

In table \ref{tabkb} we compare this way of determination of CKM
parameters with the one achieved through $K\to\pi\nu\bar\nu$ and
presented already in table 4. We set
$|V_{cb}|=0.040\pm 0.002 (\pm 0.001)$.
As can be seen in Table \ref{tabkb}, the CKM determination
using $K\to\pi\nu\bar\nu$ is competitive with the one based
on CP violation in $B$ decays in scenario I, except for $\bar\varrho$ which
is less constrained by the rare kaon processes.
On the other hand as advertised previously ${\rm Im}\lambda_t$
is better determined
in $K\to\pi\nu\bar\nu$ even if scenario II is considered.
The virtue of the comparision of the determinations
of various parameters using CP-B asymmetries with the determinations
in very clean decays $K\to\pi\nu\bar\nu$ is that any substantial deviations
from these two determinations would signal new physics beyond the
Standard Model.

\section{Final Messages}

The reduction of various scale uncertainties through the calculation
of short distance NLO-QCD corrections to a large class of weak decays and
a progress in
non-perturbative methods like HQET, HQE,
lattice calculations, chiral perturbation theory, 1/N approach
and QCD sum rules of various sorts, decreased considerably theoretical
uncertainties in the determination of the CKM matrix in the present and
future experiments. In spite of this, further progress in non-perturbative
methods is clearly desirable.

The improved data for semi-leptonic tree level B-decays combined with
HQE and HQET allowed considerable improvement in the
determination of the elements $\vcb$ and $|V_{ub}|$:
\be
\sigma(\vcb)_{97}\approx \frac{1}{3}\sigma(\vcb)_{90}
\qquad
\sigma(|V_{ub}|)_{97}\approx \frac{1}{2}\sigma(|V_{ub}|)_{90}
\ee
At present these elements are known with an accuracy of $\pm 7\%$
and $\pm 25\%$ respectively but
the hope that they will be measured one day in tree level
decays with an accuracy of $\pm 4\%$ is not fully unrealistic.

The discovery of the top quark together with the theoretical
progress mentioned above provided an improved determination
of $\vtd$. Yet the error on this element is still sizable:
roughly $\pm 25\%$.

Standard analysis of the unitarity triangle can give very useful
results provided the experimental knowledge of $\Delta M_s$,
$\vub$ and $\vcb$ and the theoretical knowledge of $\sqrt{B_B}F_B$,
$\xi$ and $B_K$ can be considerably improved. A numerical analysis
of this issue has been presented in section 3.

Useful information about $\vts$ and $\vtd$ at the level of $10-15\%$
accuracy should come from $B\to X_{d,s}\gamma$ and $B\to X_{d,s}l^+l^-$.
Moreover these decays can be efficiently used to study the physics
beyond the Standard Model.

$K^+\to\pi^+\nu\bar\nu$ should offer a measurement of $\vtd$
with an accuracy of $5-10\%$ provided its branching ratio can be
measured with an accuracy of at least $10\%$. $K_L\to\pi^0\nu\bar\nu$
appears to be the cleanest decay to study direct CP violation
and to measure $\IM\lambda_t$. With the newly proposed experiment at
BNL the present error on $\IM\lambda_t$ of roughly $30\%$ should
be reduced to $5\%$ which would be an important progress. Both
decays taken simultaneously can offer a very clean measurement
of $\sin 2\beta$ with an error of $\pm 0.05$.

The cleanest measurement of the ratio $|V_{td}/V_{ts}|$ can be
obtained from $B\to X_{d,s}\nu\bar\nu$ and to a lesser extent
from $B\to X_{d,s}l^+l^-$. However, both decays are extremly
challenging for different reasons from experimental point of view.

Clearly the future measurements of CP violation in two-body
B-decays at SLAC, KEK, Cornell, HERA-B, FNAL and LHC will open
a new chapter in the physics of weak decays and particle physics
in general irrespective of whether $\alpha$, $\beta$ and $\gamma$
can be measured very precisely. A very accurate measurement of
$\beta$ in $B\to\psi K_S$ and in $B\to\phi K_S$ appears to be
very realistic. Precise measurements of $\alpha$ and $\gamma$
seem to be much harder.
Yet one should emphasize the obvious fact that $\alpha$ and/or $\gamma$
have to be well measured in addition to $\beta$ in order
to construct the unitarity triangle on the basis of two-body
B-decays alone. In the next few years the only results here will
come from CLEO.

The most interesting will be the comparision of CKM determinations
from CP-B, $K\to\pi\nu\bar\nu$, rare B-decays and the standard
analysis of the unitarity triangle as expressed in the plot of
fig. 2. Such a comparision should give us some hints about the
physics beyond the Standard Model. The possibility of having
such a comparision within the next $5-10$ years is very exciting.

The physics around the CKM matrix has a very bright future
at least for the next 5-10 years irrespective
of whether the KM scenario for CP violation will be confirmed or
proved false in the coming experiments. By making continuos
experimental and theoretical efforts in this field we will
either achieve a high accuracy for the CKM parameters within
the Standard Model or find some hints how to generalize it.

\section*{Acknowledgments}
This was a very enjoyable and brilliantly organized symposium.
I am very grateful to the organizers, in particular to Jeff Richman and
Michael Witherell, for
these wonderful five days in Santa Barbara and for covering my local
expenses.
I would also like to thank Markus Lautenbacher for
help in producing the figures and for checking some
of my numerical calculations.
Travel support
from Max-Planck Institute for Physics in Munich is gratefully
acknowledged.

\end{document}